\documentclass[10pt,aps,twocolumn,showpacs,showkeys,tightenlines,]{revtex4}
\usepackage{amssymb}
\usepackage{bm}
\usepackage{epsfig}

\begin{document}

\title{Radiative decay of $K^-p$ system and photoproduction of $\Lambda(1405)$}

\author{Tae Keun Choi}%

\affiliation{Department of Physics, Yonsei University, Wonju,
220-710, Korea}

\author{Kyung Sik Kim and
Byung Geel Yu}%

\email{bgyu@kau.ac.kr}%
\affiliation{School of Liberal Arts \& Sciences, Korea Aerospace
University, Koyang, 412-791, Korea}



\begin{abstract}

The properties of the $\Lambda(1405)$ resonance have been
investigated from radiative decay of $K^- p\to Y\gamma$ and
photoproduction $\gamma p\to K^+\Lambda(1405)$ within the
framework of the isobar model. For a consistent result with
recently measured branching ratios, the axial vector meson
$K_1(1270)$ is taken into account. Strong and electromagnetic
coupling constants of $\Lambda(1405)$ are extracted from these
branching ratios and are applied to the analysis of
$K^+\Lambda(1405)$ photoproduction. The total and differential
cross sections are predicted.

\end{abstract}


\maketitle

\newpage
\section{Introduction}

Radiative decays $K^-p\to \Lambda\gamma$ and $K^-p\to
\Sigma^0\gamma$ are important processes to study the nature of the
$\Lambda(1405)$ resonance because of the proximity of the $K^-p$
system to the mass of the subthreshold
$\Lambda(1405)$\cite{theo1,theo2}. From parity and angular
momentum conservation, it is clear in these decay processes that
the s-channel exchange of the $\Lambda(1405)$ resonance must be
predominant at threshold\cite{isgu,thom}. Therefore, information
on the $\Lambda(1405)$ couplings can be obtained by analyzing
these radiative decays.

Measurements of the branching ratio were recently renewed by the
Brookhaven experiment. They were reported to be $R_{\Lambda\gamma}
=(0.86\pm0.07^{+0.10}_{-0.08})\times 10^{-3}$ and
$R_{\Sigma^0\gamma}=(1.44\pm0.20^{+0.12}_{-0.10})\times 10^{-3}$
\cite{exp1}. These new measurements improved previous experimental
values $R_{\Lambda\gamma}=(2.8\pm 0.8)\times 10^{-3}$ and
$R_{\Sigma^0\gamma}\leq 4\times 10^{-3}$, which might contain
attributions either from in-flight $\Lambda\pi^0$ or pile up to
the $\Lambda\gamma$ and $\Sigma^0\gamma$ events with poor energy
resolution\cite{exp2}. On the other hand, however, theoretical
estimates on these processes within the framework of the isobar
model remain untouched, with old predictions for the past
experimental values\cite{work,lowe2}.

In our previous work\cite{bgyu}, we showed that inclusion of the
t-channel exchange of the axial vector meson $K_1(1270)$  could
improve the model prediction for these ratios, because the
$K_1(1270)$ with spin parity $1^+$ was allowed by parity and
angular momentum conservation at threshold, as well as
$\Lambda(1405)$. Here, we investigate the reaction process $\gamma
p\to K^+\Lambda(1405)$ by using the result of our previous work to
constrain the process. To be more specific, we calculate cross
sections for photoproduction $\gamma p\to K^+\Lambda(1405)$ with
the coupling constants determined from the decay process of the
$K^-p$ system. Due to the scarcity of experimental data, however,
this process has rarely been studied. Therefore, the numerical
consequences in this work will be predictions for future
experiments.

In Section I, we give a brief summary of the branching ratio
$R_{Y\gamma}$ studied in Ref.\cite{bgyu} with emphasis on the
additional contribution of $K_1$ exchange, and we compare the
result with those of Refs.\cite{work,lowe2}. In the next section,
we calculate the total and differential cross sections for $\gamma
p\to K^+\Lambda(1405)$. We discuss the results in the final
section.

\section{Radiative kaon capture}

The decay width for the process $K^-(q)+p(p)\to Y(p')+\gamma(k)$
is given by
\begin{equation}\label{decay}
\Gamma_{K^-p\to Y\gamma}=|\phi_K(0)|^2\frac{M_Y |\bm{k}|}{4\pi W
m_K }\frac{1}{2}\sum_{s}\sum_{\lambda,s'}|{\cal M}|^2,
\end{equation}
where $q$, $p$, $p'$, and $k$ are the 4-momenta of the kaon,
proton, hyperon, and photon with masses $m_K$, $M$, $M_Y$,
respectively. The $\phi_K(0)$ is the wave function of the kaon
captured at the s-orbit with respect to a proton,  and $W$ is the
invariant mass of the process. The width is evaluated at threshold
$(\sqrt{s}, \bm{0})$ in the center of mass frame with the
spin-average for the initial state and the spin-sum for the final
state interacting particles. Then, the branching ratio is defined
as
\begin{equation}
R_{Y\gamma}=\frac{\Gamma_{K^-p\to Y\gamma}}{\Gamma_{K^-p\to all
}}~~,
\end{equation}
where $\Gamma_{K^-p\to all }=2W_p\, |\phi_K(0)|^2$ is the decay
width of the $K^-p$ system for all channels and $W_p=(560\pm
135)$MeV fm$^3$ is the pseudopotential of the $K^-p$
system\cite{work}.

With the transition amplitude ${\cal M}$ in Eq.(\ref{decay}) given
in Ref.\cite{work}, the t-channel $K_1(1270)$ exchange is obtained
by replacing $k\to -k$ and $q\to -q$ in the photoproduction
amplitude\cite{bgyu,bgyu1} on the basis of crossing symmetry
between the two processes. We have estimated the anomalous
coupling constant $g_{\gamma KK_1}$ by applying the vector meson
dominance to the strong coupling vertex $\rho KK_1$\cite{hagl}.
The strong coupling constants $g^{V}_{K_1pY}$, and $g^{T}_{K_1pY}$
can be determined by using the SU(3) octet
relations\cite{bgyu,durs,hep}.

In Table \ref{cb}, we list principal coupling constants for the
non-resonant Born terms and the $K^*$ and $K_1$ resonances
couplings used for the calculation of branching ratios. On the
basis of the same set of nucleon and hyperon resonances as those
of Ref.\cite{work}, the WF in Table \ref{cb} refers to the model
calculation of Ref.\cite{work}, Type I to Ref.\cite{lowe2}, and
Type II to the present work with the $K_1$ contribution to WF.
\begin{table}{}
\caption{\label{cb} Coupling constants for the non-resonant Born
terms and meson exchanges in the radiative decay processes $K^- p
\to \Lambda\gamma$ and $\Sigma^0\gamma$. Anomalous magnetic
moments are given in units of the proton magneton, $\frac{e}{2M}$,
$\kappa_p=1.793$, $\kappa_\Lambda=-0.613$,
$\kappa_{\Sigma^0}=0.619$, and $\kappa_{\Sigma^0\Lambda}=1.60$.
The coupling constants $g_{\gamma KK^*}$ and $g_{\gamma KK_1}$ are
in units of GeV$^{-1}$. Nucleon and hyperon resonances are taken
into account from Ref.\cite{work}. }
\begin{ruledtabular}\label{cb}
\begin{tabular}{cccccl}
                          &WF               & Type I     & Type II &\\
\hline\hline
$g_{Kp\Lambda}$           &-13.2&-13.2$\sqrt{0.3}$   &-13.2 &\\%
$g_{Kp\Sigma^0}$          &6.0  &6.0$\sqrt{0.3}$     &6.0 &\\%
\hline
$g_{\gamma KK^*}                    $    &0.254     &0.254  &0.254 &\\%
$g^V_{K^*p\Lambda}\,(g^T_{K^*p\Lambda})$ &-4.5(-16.7)&-4.5(-16.7)&-4.5 (-16.7)&\\%
$g^V_{K^*p\Sigma}\,(g^T_{K^*p\Sigma})$   &-2.6(3.2) &-2.6(3.2)   &-2.6 (3.2)&\\%
\hline
$g_{\gamma KK_1}                    $    &  -       &   -     &-0.6   &\\%
$g^V_{K_1p\Lambda}\,(g^T_{K_1p\Lambda})$ &  -       &   -     &-5.2 (-9.66)&\\%
$g^V_{K_1p\Sigma}\,(g^T_{K_1p\Sigma})$   &  -       &    -    &-3 (1.86) &\\%
\end{tabular}
\end{ruledtabular}
\end{table}
The coupling constants of $\Lambda(1405)$ for strong and magnetic
interactions,  $g_{Kp\Lambda_{1405}}$,
$\kappa_{\Lambda\Lambda_{1405}}$, and
$\kappa_{\Sigma^0\Lambda_{1405}}$, are not known yet. For the
calculation of the branching ratio, we use
$g_{Kp\Lambda_{1405}}=3.2$ (with an overall sign ambiguity), which
is favored in the literature. The magnetic couplings
$\kappa_{\Lambda\Lambda_{1405}}$ and
$\kappa_{\Sigma^0\Lambda_{1405}}$ are treated as parameters to be
determined from the experimental data.

\begin{figure}[t]\centering
\epsfig{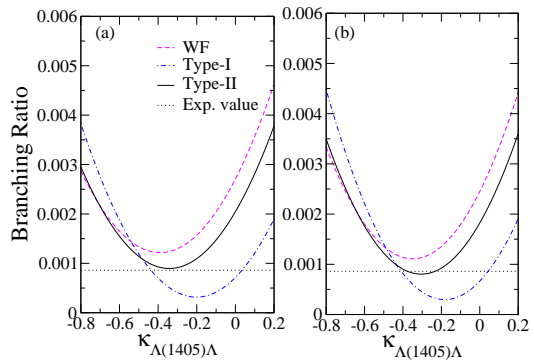} \caption[]{The branching
ratio $R_{\Lambda\gamma}$ for the $K^-p\to \Lambda\gamma$ decay
process. In each panel, the dashed line is from the WF model, the
dash-dotted line is from the Type I, and the sold line is from the
Type II. The experimental value is denoted by the horizontal
dotted line. The PS scheme is given in (a) and the PV scheme in
(b).} \label{fig:fig02}
\end{figure}

In Fig. \ref{fig:fig02}, we show the result, omitting the case of
$\Sigma^0\gamma$ decay, which does not reveal the significance of
the role of the $K_1$ because of the much smaller contribution of
the non-resonant Born terms. It is clear in Fig. \ref{fig:fig02}
that the WF model given by the dashed line can hardly account for
the experimental value in the case of $\Lambda\gamma$ decay. As
discussed in Ref.\cite{work}, the assumed value
$\kappa_{\Lambda\Lambda_{1405}}\approx -0.4$ at the minimum of the
dashed curve corresponds to $R_{\Lambda\gamma}=1.22$ for the
pseudoscalar (PS) coupling, and $1.14$ for the pseudovector (PV)
coupling scheme, neither of which is in the range of the
experimental measurement. In case of the Type I model, the
reduction of the coupling constants $g_{Kp\Lambda}\to
\sqrt{0.3}\,g_{Kp\Lambda}$ on the $R_{\Lambda\gamma}$ (and
$g_{Kp\Sigma^0}\to \sqrt{0.3}\,g_{Kp\Sigma^0}$ on the
$R_{\Sigma^0\gamma}$) leads to a suppression of the Born
contribution by a factor of 30$\%$, as claimed in
Ref.\cite{lowe2}. We find in the Type I, however, the contribution
of $\Lambda(1405)$ to the $\Lambda\gamma$ decay is by an order of
magnitude smaller than that of the WF model, or of the Type II.
Within the present framework, therefore, the dominance of
$\Lambda(1405)$ in the $\Lambda\gamma$ decay cannot be supported
by the Type I model.
The Type II model exhibits the contribution of $K_1$ meson to the
$R_{\Lambda\gamma}$ with coupling constants given in Table
\ref{cb}. These are presented as the solid lines in Fig.
\ref{fig:fig02}. The solid line in Fig. \ref{fig:fig02}(a) has an
intersection at $\kappa_{\Lambda\Lambda_{1405}}=-0.34$ to estimate
$R_{\Lambda\gamma}=0.9$ in the PS scheme while it yields a set of
value for $\kappa_{\Lambda\Lambda_{1405}}=(-0.38, -0.23)$,
consistent with $R_{\Lambda\gamma}=0.86$ for the PV scheme in Fig.
\ref{fig:fig02}(b). As shown, the deviation of the solid line from
the dashed one due to the $K_1$ contribution is significant enough
to give $\kappa_{\Lambda\Lambda_{1405}}$ a physical value.

\section{$K^+\Lambda(1405)$ photoproduction}

In this section, we investigate photoproduction $\gamma p\to
K^+\Lambda(1405)$ within the same framework of the previous
section. We restrict our aim here to a report of the numerical
result from the qualitative viewpoint, because experimental data
are very scarce for a direct comparison.

With the convention for 4-momenta,
\begin{eqnarray}\label{eq1}
\gamma(k)+ p(p)\to K^+(q)+\Lambda_{1405}(p'),
\end{eqnarray}
the cross section is given by
\begin{eqnarray}
\frac{d\sigma}{d\Omega}=\frac{M M_{\Lambda_{1405}}}{16\pi^2
W^2}\frac{|\bm q|}{|\bm k|
}\frac{1}{4}\sum_{\lambda_{\gamma},s,s'}|{\cal M}|^2\ .
\end{eqnarray}
The transition amplitude ${\cal M}$ for the photoproduction of the
negative parity $\Lambda(1405)$ is obtained by replacing
$\bar{u}_{\Lambda}(p_{\Lambda})\gamma_5\to
-\bar{u}_{\Lambda_{1405}}(p')$ in the $\gamma p\to K^+\Lambda$
photoproduction given in Ref.\cite{bgyu1}.
For consistency, we use the pseudoscalar coupling constants given
in the previous section to calculate the cross sections. Since the
reaction channel opens almost $W=1.9$ GeV above, we assume that
the roles of baryon resonances will not be significant because
fewer resonances with masses around 2 GeV are reported. For
comparison with existing calculations, we consider
$N^*(1650)\frac{1}{2}^{-}$ and $N^*(1710)\frac{1}{2}^{+}$ for the
s-channel, and $\Lambda(1405)\frac{1}{2}^{-}$ for the u-channel
exchanges. The couplings of $\Lambda(1405)$ to the vector mesons
$K^*$ and $K_1$ are currently not known. We utilize the ratio
$\frac{G^V_{K^*}}{G^V_{K_1}}\simeq -0.7$ extracted from the WJC
model fitted to the $K^+ Y$ electroproduction data\cite{wjc1}.
Resuming the ratio, we give the values for the couplings shown in
Table \ref{cc} at this exploratory stage, with assumptions that
these values be less than $g_{Kp\Lambda_{1405}}$ and that the
cross section be less than that of $K^+\Lambda$
photoproduction\cite{sinam}. We here neglect the tensor couplings
of these vector mesons for the minimal model calculation.

In Table \ref{cc}, we list the coupling constants of the present
work and compare them with those used for the decay
width\cite{work} and for the photoproduction\cite{wjc} in other
model calculations.
\begin{table}
\caption{\label{cc} Coupling constants taken from the radiative
decay process $K^- p \to Y\gamma\,{}^a $\cite{work} and
photoproduction, $\gamma p \to K^+\Lambda(1405){}^b$\cite{wjc}.
Values for the present work${}^c$ are taken from Ref.\cite{bgyu}.
Anomalous magnetic moments are given in units of the proton
magneton $\frac{e}{2M}$. $G^V_{K^*}=g_{\gamma
KK^*}g^V_{K^*p\Lambda_{1405}}$, and $G^V_{K_1}=g_{\gamma
KK_1}g^V_{K_1p\Lambda_{1405}}$. }
\begin{ruledtabular}\label{cc}
\begin{tabular}{ccccccl}
                   & WF${}^a$   & WJC${}^b$ & present work${}^c$ &\\
\hline\hline
$g_{Kp\Lambda}$    &-13.2       &4.127       &-13.2&\\%
$g_{Kp\Sigma^0}$   &6           &-0.329      &6 &\\%
\hline
$g_{K p\Lambda_{1405}}$ &3.2   &1.5 $\sim$ 3.0& 0.9 $\sim$ 3.2&\\%
$\kappa_{\Lambda_{1405}}$& -   &0.44         &-0.44  &\\%
$\kappa_{\Lambda\Lambda_{1405}}$&$\approx$ -0.4&-0.224 &-0.34&\\%
$\kappa_{\Sigma\Lambda_{1405}}$ &-0.26  &1.077& -0.23  &\\%
\hline
$g_{KN_{1710}\Lambda_{1405}}$&0.81     &0.81 &0.81    &\\%
$g_{KN_{1650}\Lambda_{1405}}$&6.5      &6.4&6.5  &\\%
$\kappa_{N_{1710}p}         $&0.03     &0.097&0.03    &\\%
$\kappa_{N_{1650}p}         $&0.32     &-0.41&-0.32    &\\%
\hline
$G^V_{K^*} $       &  -       & -        &0.276&\\%
$G^V_{K_1}$        &  -       & -        &0.395&\\%
\end{tabular}
\end{ruledtabular}
\end{table}
The values of the WJC model are taken from Ref.\cite{wjc}. In this
model, meson exchanges in the t-channel are not considered by the
duality between s- and t-channel. Thus, not only the $K^*$ but the
$K_1$ exchange is absent from the model. Note that the WF values
in Table \ref{cc} for the $K^*$ and the $K_1$ coupling constants
are irrelevant here because they are, now, coupling to the
$\Lambda(1405)$. We must address that, while there is ambiguity in
the sign of $g_{Kp\Lambda_{1405}}$, we choose the sign of the
$\kappa_{\Lambda_{1405}}$ to be opposite to the WJC one, as in
Table \ref{cc}, in consideration of the relative signs between
$g_{Kp\Lambda_{1405}}$ and $\kappa_{\Lambda_{1405}}$ in
Ref.\cite{wjc1}. We also take the sign of the coupling constant
$\kappa_{N_{1650}p}$ of WJC to be opposite to the original one in
Ref.\cite{wjc} in order for the cross section to decrease beyond
the resonance region, $E_\gamma\approx 2$ GeV. With the cutoff
$\Lambda$ for the hadron form factors taken as 1.2 GeV for the
non-resonant Born terms and 1.8 GeV for the
resonances\cite{bgyu1}, the numerical results are presented in
Fig. \ref{fig:pstcrs-1} for the total cross section and in Fig.
\ref{fig:psdcrs-1.6} for the differential cross section,
respectively.

\section{results and discussion}

\begin{figure}[tbh] \centering
\epsfig{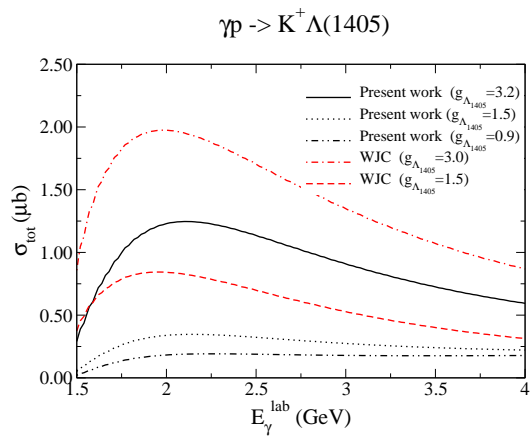} \caption{Total cross
section for the $\gamma p\to K^+\Lambda(1405)$ process. }
\label{fig:pstcrs-1}
\end{figure}

\begin{figure}[tr]
\centering \epsfig{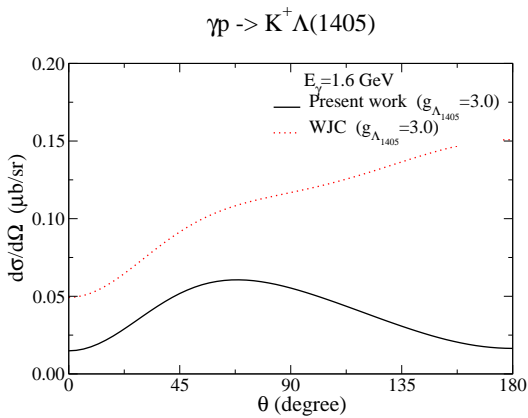}
\caption{Differential cross section near threshold
$E_{\gamma}=1.6$ GeV.} \label{fig:psdcrs-1.6}
\end{figure}

In Fig. \ref{fig:pstcrs-1}, dashed and dash-dotted curves are the
cross sections of the model with WJC coupling contants taken, each
of which results from varying $g_{Kp\Lambda_{1405}}$ from 1.5 to
3.0, respectively. The result of the present work is shown in the
dash-dot-dotted, dotted, and solid lines corresponding to the
choice of $g_{Kp\Lambda_{1405}}=0.9$, $1.5$, and $3.2$ in order.
It should be noted that all the curves in Figs. \ref{fig:pstcrs-1}
and \ref{fig:psdcrs-1.6} are calculated with the use of the same
form factors in both models, i.e., the model with WJC values and
the present work for comparison, although the former in
Ref.\cite{wjc} does not consider such form factors. In the absence
of these form factors, however, the maximum height of the WJC
cross section becomes by an order of magnitude larger than the
present one. As shown in this figure, the cross section is very
sensitive to changes in the leading coupling constant
$g_{Kp\Lambda_{1405}}$.

Figure. \ref{fig:psdcrs-1.6} shows the angular distributions in
both model calculations near threshold, $E_{\gamma}=1.6$ GeV. The
solid curve results from the present work, and the dotted one from
the model with WJC values, where $g_{Kp\Lambda_{1405}}=3.0$ is
taken in common for comparison. The angular distribution of the
former represents an apparent feature of p-wave production,
indicating a rapid increase of the t-channel kaon
exchange\cite{bgyu2}, while the latter demonstrates the backward
increase implementing the strong u-channel contribution, which in
the present study might be due to the hyperon exchanges
considered. In fact, such a distinction can be thought of as a
natural consequence of the different choice of coupling-channels;
i.e., the absence of t-channel couplings from the model with WJC
values leads to the u-channel enhancement.

In this work, we have presented an analysis of the branching ratio
of $K^-p\to Y\gamma$ and have applied the obtained coupling
constants to an analysis of the $\gamma p\to K^+\Lambda(1405)$
process in the framework of the isobar model. Introducing the
axial vector meson $K_1(1270)$ from parity and angular momentum
conservation, we have improved the model prediction for the
branching ratio. Related to this issue, we make a comment on the
unitary coupled channel approach to these processes, in which the
role of the initial state interaction in the $K^-p\to
\Lambda\gamma$ process is emphasized\cite{ramo}. In this model,
the resonance state $\Lambda(1405)$ is generated in a noble way
such as the quasi-bound state of a $\bar{K}N$ or a $\pi\Sigma$
coupled channel\cite{ramo,jkahn}. We note in Ref.\cite{ramo},
however, that such an approach to the radiative decay of the
$K^-p$ system yields an overestimate of the branching ratio, which
amounts to double the value of the measured one for the
$\Lambda\gamma$ decay; i.e., $R_{\Lambda\gamma}=1.58$ (without
cut-off $\Lambda_{\pi}$). Thus, it needs further corrections,
apart from the model dependence due to the cutoff $\Lambda_{\pi}$.

We have investigated the photoproduction $\gamma p\to
K^+\Lambda(1405)$ with relevant coupling constants constrained
from the radiative decay of the $K^-p$ system. Cross sections are
reproduced for our further understanding of the strong and the
electromagnetic properties of the $\Lambda(1405)$ through the
production mechanism. We found in this work that the magnitude of
the cross section for the $K^+\Lambda(1405)$ photoproduction was
of an order of micro barn, which was very sensitive to variations
in the leading coupling constant $g_{Kp\Lambda{405}}$. Without
resonances with masses around 2 GeV near threshold, large
sensitivity of the cross section to the change of
$g_{Kp\Lambda_{1405}}$ could be a useful tool for determining the
coupling constant by measuring cross sections in future
experiments. As discussed, the difference between the angular
distributions reproduced in both models is contrasting. They
reveal the respective features of the t-channel and the u-channel
contributions, which are to be distinguished, as well, from future
experiments.

\begin{acknowledgments}

This work was supported by the Korea Research Foundation (KRF)
Grant funded by the Korean goverment (KRF-2008-313-C00205), and by
the Korea Science and Engineering Foundation (KOSEF) Grant funded
by the Korea government(MOST)(R01-2007-000-20569-0).

\end{acknowledgments}

\end{document}